\documentclass[UTF8,11pt,a4paper,fontset=fandol]{ctexart}

\usepackage[left=2.6cm,right=2.6cm,top=2.5cm,bottom=2.5cm]{geometry}
\usepackage{amsmath,amssymb}
\usepackage{booktabs,array,multirow}
\usepackage{graphicx}
\usepackage{placeins}
\usepackage{tikz}
\usepackage{pgfplots}
\usetikzlibrary{arrows.meta,positioning,patterns}
\pgfplotsset{compat=1.18}
\usepackage{caption}
\usepackage{gbt7714}
\usepackage{enumitem}
\usepackage[hidelinks,hyperfootnotes=false]{hyperref}
\usepackage{xurl}

\setlist[enumerate]{leftmargin=2.4em,itemsep=0.2em,topsep=0.3em}
\ctexset{
  section={format=\Large\bfseries,beforeskip=1.2ex,afterskip=0.8ex},
  subsection={format=\large\bfseries,beforeskip=1.0ex,afterskip=0.5ex}
}
\xeCJKsetup{CJKecglue=\hskip 0pt}

\newenvironment{paperabstract}
  {\par\begingroup\small\setlength{\parindent}{0pt}\setlength{\parskip}{0.28em}%
   \noindent\textbf{Abstract: }\ignorespaces}
  {\par\endgroup}
\newenvironment{cnabstract}
  {\par\begingroup\small\setlength{\parindent}{0pt}\setlength{\parskip}{0.28em}%
   \noindent\textbf{摘要：}\ignorespaces}
  {\par\endgroup}

\begin{document}
\pagestyle{plain}
\renewcommand{\figurename}{Figure}
\renewcommand{\tablename}{Table}

\pagestyle{plain}

\begin{center}
  {\bfseries\LARGE A Non-Intrusive Traffic Analysis Framework for Authorization Risk Detection and Coordinated Response in Web Applications\textsuperscript{*}\par}
  \vspace{1.2em}
  {\large Siqi Lin$^{1,2,\dagger}$\quad Yuchao Luo$^{3}$\quad Zihan Zhu$^{4}$\quad Yiran Wang$^{1}$\quad Borui Qiu$^{3}$\par}
  \vspace{0.8em}
  {\small
  $^{1}$Ara Institute of Canterbury International Engineering College, Shenyang Jianzhu University, Shenyang 110168, China\par
  $^{2}$Xiamen Moefire Technology Co., Ltd., Xiamen 361000, China\par
  $^{3}$School of Computer Science and Engineering, Shenyang Jianzhu University, Shenyang 110168, China\par
  $^{4}$School of Management, Shenyang Jianzhu University, Shenyang 110168, China\par
  $^{\dagger}$Corresponding author: Siqi Lin, E-mail: \href{mailto:lsq@moefire.net}{lsq@moefire.net}\par}
\end{center}

\begingroup
\renewcommand{\thefootnote}{*}
\footnotetext{This work is one of the outcomes of the Liaoning Provincial College Student Innovation and Entrepreneurship Training Program project ``Web Application Authorization Risk Identification and Response System Based on Non-Intrusive Traffic Analysis'' (Project No. S202610153008).}
\endgroup

\begin{paperabstract}
\textbf{Objective:} Authorization violations under valid Web sessions are difficult to identify and handle in real time from traffic because they depend strongly on business semantics and exhibit few distinctive protocol-level features. This paper proposes a non-intrusive traffic analysis framework for authorization risk detection and coordinated response.

\textbf{Methods:} Request--response transactions are correlated to extract runtime context, including the access subject, business endpoint, object identifier, authentication state, and behavioral sequence. Object-access evidence, identity consistency, behavioral anomalies, authentication context, network environment, and endpoint-operation risk are mapped to interpretable risk components. Weighted fusion and high-risk priority constraints produce graded decisions that drive allow, alert, block, and external policy actions. The risk components are instantiated using deterministic and interpretable rules so that the study can evaluate multi-source evidence organization, risk fusion, and the coordinated-response loop at the framework level.

\textbf{Results:} The controlled local testbed contained 2,000 balanced labeled samples, including 1,000 normal accesses and 1,000 authorization-risk events. The framework classified 998 of the 1,000 authorization-risk events as risky and produced no false positives among normal accesses. Accuracy, precision, recall, and F1 score were 99.90\%, 100.00\%, 99.80\%, and 99.90\%, respectively. Removing runtime object evidence reduced the F1 score to 81.31\%, while removing the high-risk priority constraint reduced it to 73.90\%. In a prototype feasibility test with 100 concurrent requests and 1,000 total requests, the mean risk-decision computation latency was 0.077\,ms and the P99 latency was 0.137\,ms.

\textbf{Conclusions:} The controlled evaluation shows that the framework can organize heterogeneous runtime evidence and establish an executable authorization-risk decision and coordinated-response loop without modifying application code. The results demonstrate the feasibility of the overall mechanism in the specified validation scenarios, but do not establish general applicability in production environments.
\end{paperabstract}


\noindent\textbf{CLC number:} TP393.08

\section{Introduction}

\subsection{Background}

Web applications and API systems have become the principal service form for government and enterprise information systems, business platforms, and mobile application backends. Business-object identifiers such as user IDs, contract IDs, order IDs, file IDs, and device IDs are commonly transmitted between clients and servers as URL path segments, query parameters, form fields, or JSON request-body fields. Explicit object identifiers improve interface composability but also enlarge the attack surface for object-level authorization failures. The OWASP API Security Top 10 ranks Broken Object Level Authorization among the foremost API security risks and notes that attackers may access resources they do not own by modifying object identifiers in requests\cite{owasp-api-2023}.

Compared with unauthenticated access, SQL injection, and cross-site scripting, authorization violations depend more heavily on business semantics. Malicious requests often carry valid sessions, legitimate tokens, and correctly formatted parameters. Network-layer features, status codes, and URL blocklists alone therefore cannot determine whether the current subject is authorized to perform the requested operation. As Web systems evolve toward microservices, gateways, and separated front- and backends, access subjects, business objects, role claims, and authentication events become distributed across multiple interfaces and services, further complicating runtime authorization-risk detection.

Existing systems generally rely on code-level authorization checks, RBAC or ABAC policies, API gateway rules, and WAF signatures. These mechanisms remain the primary means of access control, but rapid interface evolution, third-party component integration, legacy-system modernization, and cross-system invocation may lead to delayed rules, invisible object semantics, or fragmented policy enforcement points. A complementary capability is therefore needed to extract runtime access evidence from request--response traffic and support independent risk detection, auditing, and coordinated response without replacing application authorization.

\subsection{Problem Analysis}

The central question in authorization-risk detection is not whether a user is logged in, but whether the current subject is authorized to access the target object or perform the requested operation. When an attacker modifies an object identifier while legitimately authenticated, or when an ordinary user invokes a privileged operation, the HTTP method, parameter format, and response status may resemble normal business traffic. Authorization violations in production may also be infrequent, slow, triggered by a single request, or distributed across endpoints, making them difficult to cover reliably with a single threshold rule.

A complementary traffic-side protection framework must address four general problems. First, it must correlate requests and responses to recover interpretable business transactions. Second, it must provide a unified representation of heterogeneous context, including subjects, endpoints, objects, authentication state, and historical behavior. Third, without direct access to the application's permission database, it must convert multi-source evidence into actionable risk decisions. Fourth, its decisions must form a response loop with gateways, WAFs, and security operations platforms.

This study focuses on the unified organization of multi-source traffic-side evidence, risk fusion, and coordinated response, examining how different evidence sources work together within the same transaction and decision chain.

\subsection{Contributions}

This paper presents a non-intrusive traffic analysis framework for authorization risk detection and coordinated response in Web applications. Its main contributions are as follows:

\begin{enumerate}[label=(\arabic*)]
  \item A request--response transaction-driven runtime context extraction process is proposed to obtain access-subject, endpoint, object, authentication-state, and behavioral-sequence information consistently at a proxy or gateway.
  \item A framework-level runtime evidence abstraction maps object-access evidence, identity consistency, behavioral anomalies, authentication context, network environment, and endpoint-operation risk into unified risk components rather than relying on a single URL rule or access-frequency threshold.
  \item An interpretable decision mechanism combines weighted fusion with high-risk priority constraints and maps the result to allow, alert, block, and external policy actions, thereby forming a decision--response loop.
  \item Overall comparison, framework-level ablation, and prototype feasibility tests are conducted in a controlled testbed to evaluate risk-decision capability, the contribution of key evidence, and basic runtime overhead.
\end{enumerate}

\section{Related Work}

\subsection{Object-Level Authorization Violations in Web Applications}

Object-level authorization violations are commonly associated with Insecure Direct Object Reference (IDOR) and Broken Object Level Authorization (BOLA). The OWASP Web Security Testing Guide defines IDOR as the direct use of user-controlled input to reference database records, files, or other resources, allowing an attacker to bypass authorization checks by modifying parameters\cite{owasp-idor}. Broken access control has also remained one of the most important Web application security risks in the OWASP Top 10\cite{owasp-top10-2025}.

Object-level authorization violations are not simple authentication failures. A user may have authenticated successfully and hold a valid token while the backend fails to verify the authorization relationship between the user and the target object. Such risks can affect read operations as well as higher-impact modification, deletion, approval, and export operations. Compared with conventional vulnerabilities, object-level authorization failures depend more strongly on business data structures and access context and are therefore harder to detect automatically.

\subsection{Rule-Based Protection and Automated Scanning}

Rule-based Web protection typically filters traffic using URL allowlists or blocklists, parameter constraints, sensitive-path matching, and signature detection. WAFs and API gateways can quickly block known attack patterns, but their policies require manual maintenance and cannot automatically understand whether user A is entitled to access object B. Frequent interface changes and inconsistent object-field names can therefore result in missing rules and false positives.

Automated black-box scanners can identify some input-validation, injection, and configuration vulnerabilities. Bau et al.\cite{bau2010} and Doup\'{e} et al.\cite{doupe2010} systematically evaluated black-box Web vulnerability scanners and reported limitations in state maintenance, post-authentication path coverage, and business-logic vulnerability detection. RESTler infers producer--consumer dependencies from API specifications and generates stateful request sequences\cite{restler2019}, improving state coverage for complex REST APIs. Its primary objective, however, is defect discovery during testing rather than online authorization decisions over production traffic. Object-level authorization testing generally requires comparisons across users, objects, and roles; generic scanners and request-sequence generation alone are insufficient to reconstruct complete authorization semantics.

\subsection{Access-Control Models and Runtime Monitoring}

Role-Based Access Control (RBAC) associates permissions with roles and reduces permission-management complexity\cite{sandhu1996}. Attribute-Based Access Control (ABAC) further incorporates subject, object, operation, and environmental attributes into authorization decisions\cite{hu2014}. These models provide a theoretical basis for internal permission design, but an external runtime traffic monitor generally cannot read complete permission configurations, database relationships, or application logic. Sun et al.\cite{sun2011} used static analysis to infer role-specific access-control assumptions, whereas AUTHSCOPE discovered vulnerable authorizations in online services through differential traffic analysis and field substitution\cite{authscope2017}. The former requires source code, and the latter targets active testing; both demonstrate the importance of cross-role or cross-subject comparison for identifying authorization defects.

Runtime monitoring methods use application state, request sequences, and historical behavior to identify anomalies. Cova et al.\cite{cova2007} proposed anomaly detection based on internal application state, while Felmetsger et al.\cite{felmetsger2010} studied runtime Web detection from the perspective of logic vulnerabilities. These studies show that business-logic security problems require state and contextual reasoning. In contrast to static analysis, active scanning, and in-application instrumentation, the present framework accepts passive or proxy-side request--response transactions as input, organizes several forms of runtime authorization-risk evidence on the traffic side, and maps risk decisions directly to graded response actions. The focus is the overall framework and coordination mechanism, while the individual evidence components are instantiated only to support the unified decision chain.

\section{Overall Framework}

\subsection{Architecture}

\begin{figure}[htbp]
  \centering
  \begin{tikzpicture}[
    x=1cm,y=1cm,
    flowbox/.style={draw=black!70,fill=black!3,rounded corners=2pt,align=center,
      minimum height=0.9cm,inner sep=5pt,font=\small,line width=0.65pt},
    keybox/.style={flowbox,fill=black!9},
    external/.style={flowbox,fill=white,dashed},
    storage/.style={flowbox,double,double distance=0.8pt,fill=black!2},
    flowline/.style={draw=black!65,line width=0.7pt},
    flowarrow/.style={flowline,-{Latex[length=2.3mm,width=1.6mm]}},
    support/.style={draw=black!55,dashed,line width=0.65pt,<->}
  ]
    \node[external,text width=2.1cm] (client) at (-4.8,0) {Client/Caller};
    \node[keybox,text width=4.2cm,font=\footnotesize] (ingress) at (0,0)
      {Non-Intrusive Traffic Ingress\\Reverse Proxy/API Gateway/Passive Tap};
    \node[external,text width=2.8cm] (protected) at (4.8,0)
      {Protected Web Application\\or API System};

    \node[flowbox,text width=4.4cm] (correlation) at (0,-1.65)
      {Request--Response\\Transaction Correlation};

    \node[flowbox,text width=2.75cm] (identity) at (-3.8,-3.35)
      {Subject and\\Authentication Context};
    \node[flowbox,text width=2.75cm] (endpoint) at (0,-3.35)
      {Endpoint and\\Object Context};
    \node[flowbox,text width=2.75cm] (behavior) at (3.8,-3.35)
      {Behavioral and\\Network Context};

    \node[flowbox,text width=3.7cm] (evidence) at (0,-5.1)
      {Runtime Authorization-Risk\\Evidence};
    \node[keybox,text width=3.7cm] (fusion) at (0,-6.65)
      {Risk Fusion and\\Priority Constraints};
    \node[flowbox,text width=3.9cm] (response) at (0,-8.2)
      {Coordinated Response\\and Audit Output};

    \node[storage,text width=3.8cm,font=\scriptsize] (state) at (4.8,-6.65)
      {Shared State Store\\Transactions and Context\\Evidence and Policies\\Audit Logs};

    \draw[flowarrow] (client) -- (ingress);
    \draw[flowarrow] (ingress) -- (protected);
    \draw[flowarrow] (ingress) -- (correlation);

    \draw[flowline] (correlation.south) -- (0,-2.45);
    \draw[flowline] (-3.8,-2.45) -- (3.8,-2.45);
    \draw[flowarrow] (-3.8,-2.45) -- (identity.north);
    \draw[flowarrow] (0,-2.45) -- (endpoint.north);
    \draw[flowarrow] (3.8,-2.45) -- (behavior.north);

    \draw[flowline] (identity.south) -- (-3.8,-4.25);
    \draw[flowline] (endpoint.south) -- (0,-4.25);
    \draw[flowline] (behavior.south) -- (3.8,-4.25);
    \draw[flowline] (-3.8,-4.25) -- (3.8,-4.25);
    \draw[flowarrow] (0,-4.25) -- (evidence.north);

    \draw[flowarrow] (evidence) -- (fusion);
    \draw[flowarrow] (fusion) -- (response);
    \draw[support] (fusion.east) -- (state.west);
  \end{tikzpicture}%
  \caption{Overall framework for authorization risk detection and coordinated response}
  \label{fig:architecture}
\end{figure}
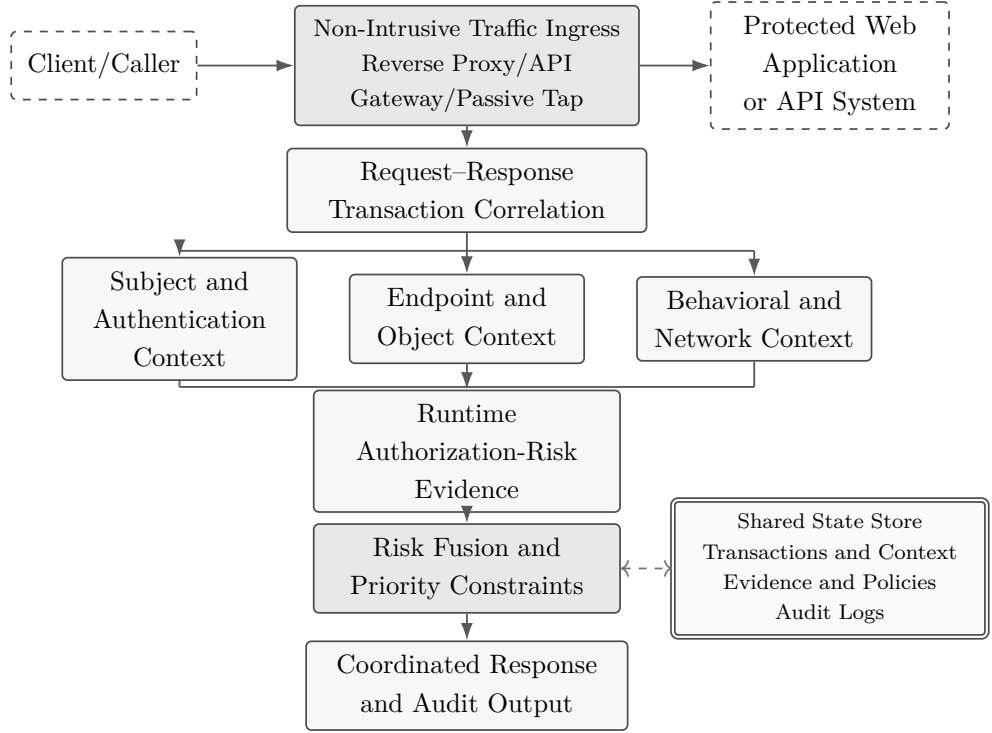

The framework consists of traffic ingress, transaction correlation, context extraction, runtime evidence organization, risk fusion, coordinated response, state management, and auditing, as shown in Fig.~\ref{fig:architecture}. It can be deployed at a reverse proxy, API gateway, or passive observation point. In online mode, the system evaluates the current request against its context and previously established runtime evidence before forwarding the request, and updates transaction state after the response returns; passive mode primarily generates audit and alert outputs. In this paper, ``non-intrusive'' means that the protected application's business code and internal authorization logic do not need to be modified; reverse-proxy and gateway deployments may still operate on the inline forwarding path.

\FloatBarrier

\subsection{Main Notation and Data Objects}

An HTTP request is the unit of online risk decision-making. After the response returns, the system forms the correlated transaction $T$ for runtime-evidence updates and auditing. Table~\ref{tab:notation} lists the main data objects used by the framework. UserKey, EndpointKey, and ObjectId identify the access subject, normalized endpoint, and business-resource instance, respectively. The set $\mathcal{E}_{u,r}(t)$ contains the runtime object-access evidence established for subject $u$ and resource type $r$ before time $t$.

\begin{table}[htbp]
  \centering
  \caption{Main notation and meanings}
  \label{tab:notation}
  \begin{tabular}{p{3.0cm}p{10.2cm}}
    \toprule
    Symbol & Meaning \\
    \midrule
    $T$ & A correlated HTTP request--response transaction and its temporal, network, and business context, used for evidence updates and auditing \\
    UserKey & Access-subject identifier derived from an account, session, or low-confidence temporary identity \\
    EndpointKey & Endpoint identifier formed by the HTTP method and normalized path \\
    ObjectId & Candidate identifier of a business-resource instance \\
    $\mathcal{E}_{u,r}(t)$ & Basic runtime object-access evidence established for subject $u$ and resource type $r$ before time $t$ \\
    \bottomrule
  \end{tabular}
\end{table}

\subsection{Data Processing and Deployment Modes}

The system first receives an HTTP/HTTPS request sent by a client to a Web application. It extracts context such as UserKey, EndpointKey, ObjectId, and authentication state from request headers, cookies, JWTs, query parameters, form parameters, and JSON request bodies, and combines this context with existing object evidence and recent access sequences to form the risk components. Before forwarding the request, the risk-fusion component outputs an overall risk level, according to which the response component allows, alerts, blocks, or sends instructions to an external policy enforcement point. The decision for the current request uses only the context available at request arrival and runtime evidence established by earlier transactions. For an allowed request, after the response returns, the system correlates the request and response as transaction $T$, adds the response summary and business outcome, and updates evidence for subsequent requests and the audit state; the current response is not used retrospectively to classify the same request.

In reverse-proxy and API-gateway deployments, the system can obtain complete transactions from plaintext HTTP after TLS termination and support online blocking. In passive mode, it observes mirrored traffic, gateway logs, or audit logs and produces alerts. When application-layer plaintext is unavailable, only metadata-level analysis is possible, and complete extraction of object identifiers and business outcomes cannot be guaranteed.

\subsection{Prototype Implementation and Study Scope}

This study focuses on the overall framework and the coordination among its components. To make the framework-level mechanism executable and reproducible, the prototype instantiates each risk component with deterministic and interpretable rules. Object-access evidence is maintained as a binary set confirmed jointly by a valid subject, a successful business outcome, preconfigured transaction semantics, and the current risk decision. Identity and authentication risk is calculated from consistency among account, session, token, role, and recent authentication context. Behavioral risk is determined from object switching, request frequency, and sequence characteristics within a time window. Endpoint-operation risk is configured according to the HTTP method, path semantics, and predefined business impact level.

The evaluation examines whether these components can operate within a unified data flow and decision chain, whether the principal evidence and fusion mechanisms affect the decision outcome, and whether the prototype can perform online decisions in the controlled testbed. The conclusions are therefore limited to framework-level mechanism validation under the stated experimental conditions.

\section{Key Mechanisms}

\subsection{Request--Response Correlation and Context Extraction}

Request--response correlation restores protocol messages as interpretable business transactions. Transaction $T$ includes the request method, normalized path, headers, parameter and request-body summaries, response status, response-body summary, business-success indicator, client address, and timestamp. HTTP semantics and URI structure follow RFC 9110\cite{rfc9110}, and JWT claim fields follow RFC 7519\cite{rfc7519}.

Proxy and gateway deployments can correlate messages deterministically using connection context, request sequence numbers, TraceId values, or gateway-injected identifiers. Passive deployments can approximate correlation using the five-tuple, request features, and a time window. For persistent connections, asynchronous interfaces, and batch requests, the system records correlation confidence and avoids using incomplete transactions to update high-confidence evidence.

UserKey preferentially uses an account identifier, followed by a session or token identifier. If these values are unavailable, a low-confidence temporary identifier can be derived from the source address, User-Agent, and device characteristics. EndpointKey combines the HTTP method with a normalized path in which numeric strings, UUIDs, and random identifiers are replaced by placeholders. ObjectId is extracted from paths, query parameters, forms, or JSON fields; field semantics, value patterns, and response structure are used to exclude obvious page numbers, timestamps, and monetary values.

\subsection{Runtime Authorization-Risk Evidence}

The framework maps heterogeneous context to six basic risk components:

\begin{enumerate}[label=(\arabic*)]
  \item Object-access evidence risk $x_{\mathrm{obj}}$, indicating whether the current ObjectId has confirmed runtime access evidence.
  \item Identity-consistency risk $x_{\mathrm{id}}$, indicating obvious conflicts among account, session, token, and role claims.
  \item Behavioral-anomaly risk $x_{\mathrm{beh}}$, indicating rapid object switching, sequential identifier access, or anomalous request sequences.
  \item Authentication-context risk $x_{\mathrm{auth}}$, indicating whether a sensitive operation is supported by a recent valid authentication or privilege-change event.
  \item Network-environment risk $x_{\mathrm{net}}$, representing auxiliary evidence such as address drift, anomalous proxies, or shared egress.
  \item Endpoint-operation risk $s_{\mathrm{ep}}$, representing the potential business impact of the current HTTP method and endpoint operation.
\end{enumerate}

The basic object-access evidence is defined in Eq.~(\ref{eq:evidence-set}). In the prototype, a transaction is treated as trusted only when the subject is valid, the response confirms business success, the transaction matches preconfigured confirmation semantics, and the request has not been assigned a medium- or high-risk decision. Such a transaction may update evidence for later requests but cannot alter the decision already made for itself. This definition provides an executable evidence interface for framework-level validation.

\begin{equation}
  \mathcal{E}_{u,r}(t)=
  \left\{o\mid o\text{ is confirmed for subject }u
  \text{ by a trusted transaction before }t\right\}
  \label{eq:evidence-set}
\end{equation}

If the current object is already present in $\mathcal{E}_{u,r}(t)$, its object-evidence risk is low. If the object is absent and is accompanied by anomalous object switching, a sensitive operation, or an identity-context conflict, the risk increases. To avoid penalizing users on shared or mobile networks, network-environment risk alone never triggers denial.

\subsection{Risk Fusion and High-Risk Priority Constraints}

All risk components are normalized to $[0,1]$. The base risk is a fixed-weight sum with $w_i\geq0$ and $\sum_{i=1}^{6}w_i=1$. To prevent several low-risk components from masking a small amount of high-severity evidence, the framework applies interpretable high-risk priority constraints. The total risk is calculated as

\begin{equation}
\begin{aligned}
  R_{\mathrm{base}}&=
  \min\!\left\{1,\max\!\left[0,\sum_{i=1}^{6}w_i x_i\right]\right\},\\
  R_{\mathrm{total}}&=
  \max\!\left(R_{\mathrm{base}},
  \max\!\left(\{h_j\mid j\in\mathcal{H}\}\cup\{0\}\right)\right),
\end{aligned}
\label{eq:risk-total}
\end{equation}

where $\boldsymbol{x}=(x_{\mathrm{obj}},x_{\mathrm{id}},x_{\mathrm{beh}},x_{\mathrm{auth}},x_{\mathrm{net}},s_{\mathrm{ep}})$; $\mathcal{H}$ is the set of triggered high-risk conditions, and $h_j\in[0,1]$ is the corresponding lower risk bound. Including $0$ in the inner maximum defines the result when no high-risk condition is triggered. A high-risk condition is triggered only by an interpretable combination of evidence, such as ``missing object evidence and a sensitive operation'' or ``conflicting identity claims without a valid authentication event,'' rather than by a single network feature.

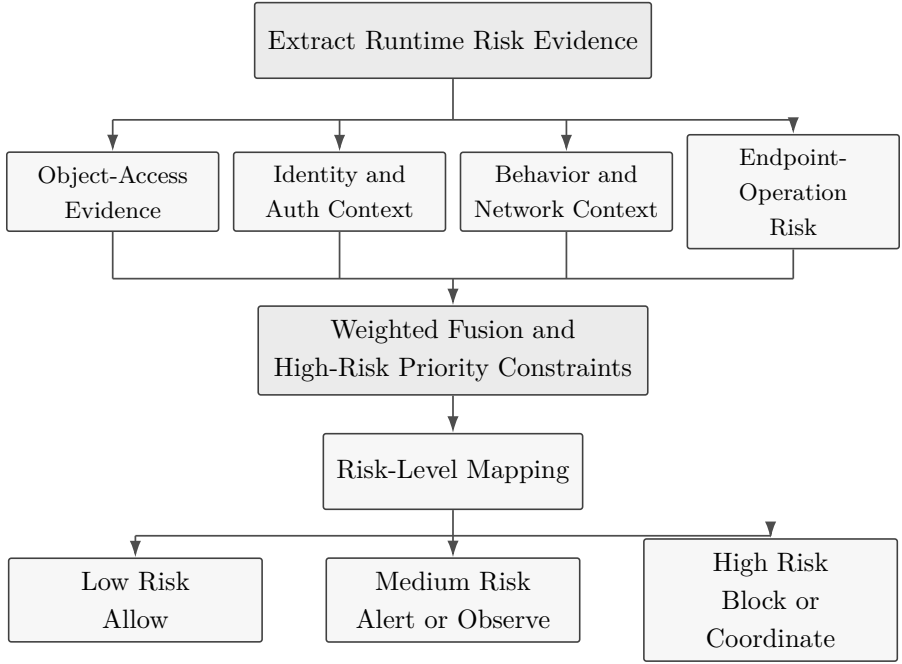
\begin{figure}[htbp]
  \centering
  \begin{tikzpicture}[
    x=1cm,y=1cm,
    flowbox/.style={draw=black!75,fill=black!3,rounded corners=1pt,align=center,
      minimum height=1.0cm,inner sep=5pt,font=\small,line width=0.6pt},
    keybox/.style={flowbox,fill=black!8},
    flowline/.style={draw=black!70,line width=0.65pt},
    flowarrow/.style={flowline,-{Latex[length=2.2mm,width=1.5mm]}}
  ]
    \node[keybox,minimum width=3.8cm] (extract) at (0,0) {Extract Runtime Risk Evidence};

    \node[flowbox,text width=2.45cm,font=\footnotesize] (object) at (-4.5,-2.0)
      {Object-Access\\Evidence};
    \node[flowbox,text width=2.45cm,font=\footnotesize] (identity) at (-1.5,-2.0)
      {Identity and\\Auth Context};
    \node[flowbox,text width=2.45cm,font=\footnotesize] (behavior) at (1.5,-2.0)
      {Behavior and\\Network Context};
    \node[flowbox,text width=2.45cm,font=\footnotesize] (endpoint) at (4.5,-2.0)
      {Endpoint-Operation\\Risk};

    \node[keybox,text width=4.8cm] (fusion) at (0,-4.1)
      {Weighted Fusion and\\High-Risk Priority Constraints};
    \node[flowbox,minimum width=3.2cm] (threshold) at (0,-5.7) {Risk-Level Mapping};

    \node[flowbox,text width=3.0cm] (low) at (-4.2,-7.4) {Low Risk\\Allow};
    \node[flowbox,text width=3.0cm] (medium) at (0,-7.4) {Medium Risk\\Alert or Observe};
    \node[flowbox,text width=3.0cm] (high) at (4.2,-7.4) {High Risk\\Block or\\Coordinate};

    \draw[flowline] (extract.south) -- (0,-1.05);
    \draw[flowline] (-4.5,-1.05) -- (4.5,-1.05);
    \draw[flowarrow] (-4.5,-1.05) -- (object.north);
    \draw[flowarrow] (-1.5,-1.05) -- (identity.north);
    \draw[flowarrow] (1.5,-1.05) -- (behavior.north);
    \draw[flowarrow] (4.5,-1.05) -- (endpoint.north);
    \draw[flowline] (object.south) -- (-4.5,-3.15);
    \draw[flowline] (identity.south) -- (-1.5,-3.15);
    \draw[flowline] (behavior.south) -- (1.5,-3.15);
    \draw[flowline] (endpoint.south) -- (4.5,-3.15);
    \draw[flowline] (-4.5,-3.15) -- (4.5,-3.15);
    \draw[flowarrow] (0,-3.15) -- (fusion.north);
    \draw[flowarrow] (fusion) -- (threshold);
    \draw[flowline] (threshold.south) -- (0,-6.55);
    \draw[flowline] (-4.2,-6.55) -- (4.2,-6.55);
    \draw[flowarrow] (-4.2,-6.55) -- (low.north);
    \draw[flowarrow] (0,-6.55) -- (medium.north);
    \draw[flowarrow] (4.2,-6.55) -- (high.north);
  \end{tikzpicture}%
  \caption{Runtime risk-evidence fusion and response process}
  \label{fig:risk-fusion}
\end{figure}

The decision process is shown in Fig.~\ref{fig:risk-fusion}.

\subsection{Coordinated Response}

The response component divides requests into low-, medium-, and high-risk classes according to $R_{\mathrm{total}}$ and policy thresholds. Low-risk requests are allowed and recorded in the audit log. Medium-risk requests trigger alerts, session flags, or observation policies. In online deployment, high-risk requests are blocked, and policy instructions can be sent to a WAF, API gateway, zero-trust gateway, or security operations platform.

The framework retains fail-open bypass, policy rollback, and allowlist mechanisms. If detection becomes unavailable or the context is incomplete, the system can degrade to observation mode so that a security-component failure does not directly interrupt the application. All evidence, decisions, and response actions are written to audit records for subsequent explanation and policy review.

\section{Experimental Design and Results}

\subsection{Experimental Environment and Data}

To evaluate the feasibility of the overall mechanism, a Web application testbed and a reverse-proxy detection prototype were constructed in a controlled local environment. Experimental scripts generated reproducible request sequences and recorded the ground-truth label, framework decision, and decision latency for each request. The experiment contained no real user data, and its sample distribution does not represent the natural prevalence of authorization-risk events in production.

The dataset contained 2,000 samples: 1,000 normal accesses, 600 object-access risk events, and 400 permission-and-context risk events, as shown in Table~\ref{tab:dataset-composition}. Object-access risk events consisted of horizontal object substitution and sequential object probing. Permission-and-context risk events consisted of privileged operations, conflicting identity claims, and anomalous authentication context. These aggregate categories evaluate how different evidence sources work together within the same framework; they are not used for separate fine-grained performance comparisons.

\begin{table}[htbp]
  \centering
  \caption{Experimental sample composition}
  \label{tab:dataset-composition}
  \begin{tabular}{lrr}
    \toprule
    Sample category & Count & Percentage/\% \\
    \midrule
    Normal access & 1000 & 50.0 \\
    Object-access risk event & 600 & 30.0 \\
    Permission-and-context risk event & 400 & 20.0 \\
    \midrule
    Total & 2000 & 100.0 \\
    \bottomrule
  \end{tabular}
\end{table}

\subsection{Baselines and Evaluation Metrics}

Four methods were compared:

\begin{enumerate}[label=(\arabic*)]
  \item URL rules, which classify requests using path terms such as admin, delete, role, export, and download.
  \item Frequency detection, which identifies anomalies using the number of requests or distinct ObjectId changes within a time interval.
  \item Rules plus frequency, which takes the union of the first two strategies.
  \item The proposed framework, which jointly uses transaction correlation, object-access evidence, identity and authentication context, behavioral anomalies, network environment, and endpoint-operation risk.
\end{enumerate}

The evaluation metrics are accuracy, precision, recall, F1 score, false-positive rate (FPR), and false-negative rate (FNR). TP denotes an authorization-risk event correctly classified as risky; FP denotes a normal access misclassified as risky; TN denotes a normal access correctly classified as non-risky; and FN denotes an authorization-risk event not classified as risky. The metrics are calculated as

\begin{equation}
\begin{aligned}
  \mathrm{Accuracy}&=\frac{TP+TN}{TP+FP+TN+FN}, &
  \mathrm{Precision}&=\frac{TP}{TP+FP},\\
  \mathrm{Recall}&=\frac{TP}{TP+FN}, &
  F1&=\frac{2\,\mathrm{Precision}\,\mathrm{Recall}}
  {\mathrm{Precision}+\mathrm{Recall}},\\
  \mathrm{FPR}&=\frac{FP}{FP+TN}, &
  \mathrm{FNR}&=\frac{FN}{FN+TP}.
\end{aligned}
\label{eq:metrics}
\end{equation}

\subsection{Framework Validation Results}

Table~\ref{tab:overall-results} reports the validation results of the four methods on 2,000 samples. URL rules classified 517 authorization-risk events as risky but also misclassified 231 normal accesses, yielding an F1 score of 59.15\%. Frequency detection produced no false positives among normal accesses but classified only 552 authorization-risk events as risky, for a recall of 55.20\%. Rules plus frequency improved recall to 83.70\% but still misclassified 231 normal accesses. The proposed framework classified 998 authorization-risk events as risky and produced no false positives among normal accesses. Its accuracy, recall, and F1 score were 99.90\%, 99.80\%, and 99.90\%, respectively.

\begin{table}[htbp]
  \centering
  \caption{Comparison of overall framework validation results}
  \label{tab:overall-results}
  \resizebox{\textwidth}{!}{%
  \begin{tabular}{lrrrrrrrr}
    \toprule
    Method & TP & FP & TN & FN & Accuracy/\% & Precision/\% & Recall/\% & F1/\% \\
    \midrule
    URL rules & 517 & 231 & 769 & 483 & 64.30 & 69.12 & 51.70 & 59.15 \\
    Frequency detection & 552 & 0 & 1000 & 448 & 77.60 & 100.00 & 55.20 & 71.13 \\
    Rules + frequency & 837 & 231 & 769 & 163 & 80.30 & 78.37 & 83.70 & 80.95 \\
    Proposed framework & 998 & 0 & 1000 & 2 & 99.90 & 100.00 & 99.80 & 99.90 \\
    \bottomrule
  \end{tabular}}
\end{table}

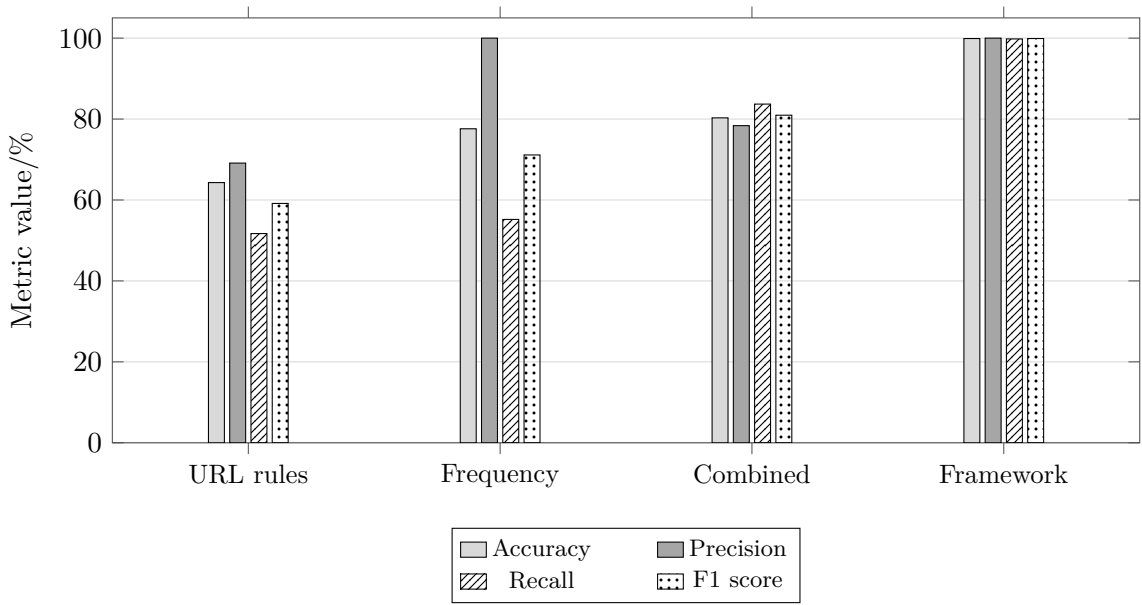
\begin{figure}[htbp]
  \centering
  \begin{tikzpicture}
    \begin{axis}[
      width=0.96\textwidth,height=7.2cm,
      ybar,bar width=6pt,
      ymin=0,ymax=105,
      ylabel={Metric value/\%},
      symbolic x coords={URL rules,Frequency,Combined,Framework},
      xtick=data,
      x tick label style={font=\small,align=center},
      ymajorgrids=true,grid style={black!15},
      legend style={at={(0.5,-0.20)},anchor=north,legend columns=2,font=\footnotesize,
        /tikz/every even column/.append style={column sep=0.8cm}},
      legend image code/.code={\draw[#1] (0cm,-0.10cm) rectangle (0.34cm,0.10cm);},
      enlarge x limits=0.18,
      axis line style={black!70},tick style={black!70}
    ]
      \addplot[draw=black,fill=black!15] coordinates {(URL rules,64.30) (Frequency,77.60) (Combined,80.30) (Framework,99.90)};
      \addplot[draw=black,fill=black!35] coordinates {(URL rules,69.12) (Frequency,100.00) (Combined,78.37) (Framework,100.00)};
      \addplot[draw=black,pattern=north east lines,pattern color=black] coordinates {(URL rules,51.70) (Frequency,55.20) (Combined,83.70) (Framework,99.80)};
      \addplot[draw=black,pattern=dots,pattern color=black] coordinates {(URL rules,59.15) (Frequency,71.13) (Combined,80.95) (Framework,99.90)};
      \legend{Accuracy,Precision,Recall,F1 score}
    \end{axis}
  \end{tikzpicture}
  \caption{Comparison of validation metrics across methods}
  \label{fig:overall-metrics}
\end{figure}

Figure~\ref{fig:overall-metrics} provides a visual comparison of the four evaluation metrics across the methods.

In the controlled samples, URL keywords could not represent changes in object ownership, while frequency thresholds failed to cover low-frequency authorization-risk events. The proposed framework's results arose from the joint use of multiple forms of context within the same transaction and risk model. Because a fixed testbed script generated the data, the metrics demonstrate only the feasibility of framework-level decisions in the specified scenarios and should not be generalized as universal production performance.

\subsection{Framework-Level Ablation Study}

To evaluate the contribution of the key mechanisms, three framework-level ablations were selected to isolate the effects of key evidence categories and the priority-constraint mechanism. The results are listed in Table~\ref{tab:ablation}. Removing runtime object-access evidence reduced recall from 99.80\% to 68.50\%, showing that object context is an important basis for authorization-risk identification. Removing behavioral-sequence evidence yielded an F1 score of 99.70\%, indicating that this evidence mainly supplements sequential object switching and slow probing. Removing the high-risk priority constraint reduced recall to 58.60\%, suggesting that linear weighting alone can allow low-risk components to mask high-severity evidence.

\begin{table}[htbp]
  \centering
  \caption{Framework-level ablation results}
  \label{tab:ablation}
  \begin{tabular}{lrrrr}
    \toprule
    Ablation & Accuracy/\% & Recall/\% & F1/\% & FNR/\% \\
    \midrule
    Full framework & 99.90 & 99.80 & 99.90 & 0.20 \\
    Without object-access evidence & 84.25 & 68.50 & 81.31 & 31.50 \\
    Without behavioral-sequence evidence & 99.70 & 99.40 & 99.70 & 0.60 \\
    Without high-risk priority constraints & 79.30 & 58.60 & 73.90 & 41.40 \\
    \bottomrule
  \end{tabular}
\end{table}

Here, object-access evidence refers to the basic binary set in Eq.~(\ref{eq:evidence-set}). The ablation result establishes the contribution of this evidence category within the current framework-level implementation.

\subsection{Prototype Runtime Feasibility}

To confirm that the framework can operate in a proxy path, a single-host prototype test was conducted with 100 concurrent requests and 1,000 total requests. The results are shown in Table~\ref{tab:performance}. The mean decision-computation latency was 0.077\,ms, the P99 latency was 0.137\,ms, and the end-to-end throughput was 91.924\,RPS. Mean computation latency includes only risk-evidence retrieval and decision processing, whereas throughput also includes proxy forwarding, testbed application processing, client scheduling, and queueing; the two measurements are not directly convertible.

\begin{table}[htbp]
  \centering
  \caption{Single-host prototype runtime feasibility test}
  \label{tab:performance}
  \resizebox{\textwidth}{!}{%
  \begin{tabular}{rrrrrr}
    \toprule
    Concurrency & Requests & Throughput/RPS & Mean decision latency/ms & P99/ms & Peak memory/MB \\
    \midrule
    100 & 1000 & 91.924 & 0.077 & 0.137 & 357.426 \\
    \bottomrule
  \end{tabular}}
\end{table}

This result shows only that the current single-host prototype can complete online risk decisions. It does not support conclusions about performance at other operating scales or under long-duration and complex deployment conditions.

\subsection{Summary of Experimental Findings}

In the constructed scenarios, the multi-source runtime-evidence framework achieved higher recall and F1 scores than the selected URL-rule and frequency-based baselines. The framework-level ablation shows that object-access evidence and high-risk priority constraints are the principal mechanisms affecting recall in the current prototype. The single-host test verifies basic feasibility for online risk decisions. Together, these results support the mechanism-level feasibility of the overall framework in the specified validation environment.

\section{Discussion}

\subsection{Applicability and Operating Mechanism}

This study provides unified definitions of data objects, evidence interfaces, risk constraints, and response processes to explain how multi-source traffic-side evidence forms actionable authorization-risk decisions. Framework validation and ablation show that object-access evidence, high-risk priority constraints, and the joint use of identity, behavior, and endpoint context affect the decision outcome. The overall design therefore forms a relatively complete detection and response loop in the specified controlled environment.

The framework is applicable to Web applications and API systems that require enhanced runtime authorization-risk auditing. In new systems, it can operate as a monitoring layer complementary to RBAC or ABAC. In legacy systems, it can provide independent auditing without large-scale application-code changes. In multi-gateway or multi-microservice environments, it can organize cross-endpoint context at a common ingress. The framework can also cooperate with continuous evaluation, least privilege, and policy enforcement points in a zero-trust architecture\cite{rose2020}, exporting risk results to gateways or security operations platforms for unified response.

In the current prototype, $\mathcal{E}_{u,r}(t)$ is represented as a binary object-evidence set, identity and authentication context is evaluated through deterministic consistency rules, and endpoint risk is configured from request semantics and predefined business impact. The conclusions therefore concern the feasibility of the unified evidence-organization, risk-fusion, and coordinated-response mechanism.

\subsection{Threats to Validity and Limitations}

The experiment has four principal threats to validity. First, the samples were generated by local testbed scripts and were balanced between positive and negative cases; this differs from production environments, where authorization-risk events are rare and business traffic has a long-tail distribution. Second, the baselines consist mainly of URL rules and frequency strategies and do not cover all static-analysis, differential-testing, and learning-based methods. Third, evidence-confirmation rules, risk weights, and decision thresholds depend on the semantics of the testbed interfaces and would require recalibration for other systems. Fourth, the performance measurements come from a single-host prototype and do not reflect long-duration operation or complex deployment conditions. The findings should therefore be interpreted as framework validation in a controlled environment rather than universal performance for arbitrary Web applications.

The method has three general limitations. First, traffic-side analysis depends on the visibility of application-layer content and key business fields; when encrypted content is unavailable or identifiers are heavily obfuscated, evidence completeness may be reduced. Second, the prototype's evidence rules, risk weights, thresholds, and endpoint configuration are tied to the semantics of the testbed and require recalibration when transferred to another application. Third, the controlled balanced samples and single-host implementation do not capture the class imbalance, long-term state variation, and deployment complexity of production traffic. Further evaluation with heterogeneous Web applications and real gateway traffic is needed to assess cross-scenario applicability and operational stability.

\section{Conclusion}

This paper addresses the difficulty of identifying and responding to authorization risks under valid Web sessions from traffic by proposing an overall non-intrusive traffic analysis framework. The framework treats each HTTP request as the unit of online risk decision-making and uses the correlated request--response transaction to organize and update object-access evidence, identity and authentication context, behavioral sequences, network environment, and endpoint-operation risk. Weighted fusion and high-risk priority constraints then produce graded response decisions.

On a controlled testbed with synthetic balanced samples, the framework achieved higher risk-decision metrics than the selected URL-rule, frequency-detection, and combined baselines. Framework-level ablation confirmed the roles of object-access evidence and high-risk priority constraints, and the single-host test showed that the prototype could complete online risk decisions. These results demonstrate mechanism-level feasibility in the specified validation environment; broader applicability requires evaluation with heterogeneous Web applications and real gateway traffic.

\section*{Author Contributions}

Siqi Lin: conceptualization, overall framework design, primary drafting, and final revision.

Yuchao Luo and Borui Qiu: system design and implementation, experimental-scenario construction, and authorization-risk sample design.

Yiran Wang: experimental data organization, result analysis, and figure preparation.

Zihan Zhu: related-work review, manuscript formatting, and reference checking.

\clearpage
\phantomsection
\begin{center}
  {\heiti\zihao{2}\bfseries 中文版本\par}
\end{center}
\vspace{1em}

\setcounter{section}{0}
\setcounter{subsection}{0}
\setcounter{figure}{0}
\setcounter{table}{0}
\setcounter{equation}{0}
\renewcommand{\theHsection}{zh.\arabic{section}}
\renewcommand{\theHsubsection}{zh.\arabic{section}.\arabic{subsection}}
\renewcommand{\theHfigure}{zh.\arabic{figure}}
\renewcommand{\theHtable}{zh.\arabic{table}}
\renewcommand{\theHequation}{zh.\arabic{equation}}
\renewcommand{\figurename}{图}
\renewcommand{\tablename}{表}
\ctexset{
  section={format=\heiti\zihao{4},beforeskip=1.2ex,afterskip=0.8ex},
  subsection={format=\heiti\zihao{-4},beforeskip=1.0ex,afterskip=0.5ex}
}

\pagestyle{plain}

\begin{center}
  {\heiti\zihao{2}\bfseries 基于非侵入式流量分析的Web应用\\越权风险识别与联动处置框架\textsuperscript{*}\par}
  \vspace{1.2em}
  {\zihao{4}林思齐$^{1,2,\dagger}$\quad 罗宇超$^{3}$\quad 朱梓涵$^{4}$\quad 王怡然$^{1}$\quad 邱博睿$^{3}$\par}
  \vspace{0.8em}
  {\zihao{-5}
  $^{1}$（沈阳建筑大学中新国际工程学院，沈阳 110168）\par
  $^{2}$（厦门萌火科技有限公司，厦门 361000）\par
  $^{3}$（沈阳建筑大学计算机科学与工程学院，沈阳 110168）\par
  $^{4}$（沈阳建筑大学管理学院，沈阳 110168）\par
  $^{\dagger}$通信作者：林思齐，E-mail: \href{mailto:lsq@moefire.net}{lsq@moefire.net}\par}
\end{center}

\begingroup
\renewcommand{\thefootnote}{*}
\footnotetext{本文系辽宁省大学生创新创业训练计划基金项目“基于非侵入式流量分析的Web应用越权访问风险识别与处置系统”（项目编号：S202610153008）的研究成果之一。}
\endgroup

\begin{cnabstract}
\textbf{【目的】}针对Web应用合法会话下的越权访问具有业务语义依赖强、协议层特征不显著且难以在流量侧实时处置的问题，提出一种基于非侵入式流量分析的越权风险识别与联动处置框架。

\textbf{【方法】}通过请求--响应事务关联提取访问主体、业务端点、对象标识、认证状态和行为序列等运行时上下文，将对象访问证据、身份一致性、行为异常、认证上下文、网络环境和端点操作风险统一映射为可解释的风险分量，并采用加权融合与高风险优先约束形成分级判定结果，驱动放行、告警、阻断和外部策略联动。各风险分量采用确定性、可解释的规则化方法实现，本文重点验证多源证据组织、风险融合和联动处置闭环的总体机制。

\textbf{【结果】}在实验环境中构造的2000条类别平衡样本中，正常访问样本和授权风险事件样本各1000条。框架验证结果显示，1000条授权风险事件中有998条被判定为风险，正常访问未出现误判；准确率、精确率、召回率和F1值分别为99.90\%、100.00\%、99.80\%和99.90\%。去除运行时对象证据后F1值降至81.31\%，去除高风险优先约束后降至73.90\%。在并发数为100、请求总数为1000的运行可行性测试中，单次风险决策的平均计算时延为0.077\,ms，P99时延为0.137\,ms。

\textbf{【结论】}实验结果表明，该框架能够在不修改业务代码的条件下组织多源运行时证据，并形成可执行的越权风险判定与联动处置闭环。上述结果仅表明总体机制在所设验证场景下具有可行性，尚不足以说明其在生产环境中的普遍适用性。
\end{cnabstract}

\noindent\textbf{关键词：}Web应用安全；越权访问；非侵入式流量分析；运行时证据；联动处置

\noindent\textbf{中图分类号：}TP393.08

\section{引言}

\subsection{研究背景}

Web应用与API系统已经成为政企信息系统、业务中台和移动应用后端的主要服务形态。用户编号、合同编号、订单编号、文件编号和设备编号等业务对象标识，通常以URL路径段、查询参数、表单字段或JSON请求体字段的形式在客户端与服务端之间传递。对象标识的显式传递提高了接口调用与系统集成效率，也扩大了对象级授权失效的攻击面。OWASP API Security Top 10将Broken Object Level Authorization列为API安全的首要风险之一，指出攻击者可能通过修改请求中的对象标识访问不属于自己的资源\cite{owasp-api-2023}。

与未登录访问、SQL注入和跨站脚本等传统漏洞相比，越权访问具有更强的业务语义依赖。攻击请求往往携带有效会话、有效令牌和正常参数格式，仅依靠网络层特征、状态码或URL黑名单难以判断当前主体是否有权执行目标操作。随着Web系统向微服务化、网关化和前后端分离演进，访问主体、业务对象、角色声明和认证事件分布在多个接口与服务之间，授权风险的运行时识别难度进一步上升。

现有系统通常依赖代码级鉴权、RBAC/ABAC策略、API网关规则和WAF签名完成防护。这些机制仍是访问控制的主体，但在接口快速迭代、第三方组件接入、存量系统改造和跨系统调用等场景中，可能出现规则滞后、对象语义不可见或策略执行点分散的问题。因此，需要在不替代业务鉴权的前提下，从请求--响应流量中提取运行时访问证据，形成独立的风险识别、审计和联动处置能力。

\subsection{问题分析}

越权风险的核心不是“用户是否已登录”，而是“当前访问主体是否有权访问目标对象或执行当前操作”。当攻击者在合法登录状态下修改对象编号，或普通用户尝试调用高权限操作时，请求的HTTP方法、参数格式和响应状态均可能与正常业务请求相似。生产环境中的越权行为还具有低频、慢速、单次触发和跨端点分布等特点，难以由单一阈值规则稳定覆盖。

从流量侧建立补充防护框架，需要解决四个共性问题：一是关联请求与响应，恢复可解释的业务事务；二是统一表示访问主体、端点、对象、认证状态和历史行为等异构上下文；三是在缺乏业务权限数据库的情况下，将多源证据转换为可执行的风险判断；四是使风险判定结果能够与网关、WAF和安全运营平台形成处置闭环。

本文聚焦流量侧多源证据的统一组织、风险融合与联动处置，研究不同证据来源如何在同一事务和同一决策链中协同发挥作用。

\subsection{本文贡献}

本文提出一种基于非侵入式流量分析的Web应用越权风险识别与联动处置框架，主要贡献如下：

（1）提出请求--响应事务驱动的运行时上下文提取流程，在代理或网关侧统一获得访问主体、端点、对象、认证状态和行为序列等信息；

（2）构建统一的运行时证据抽象，将对象访问证据、身份一致性、行为异常、认证上下文、网络环境和端点操作风险映射为统一风险分量，而不依赖单一URL规则或访问频率；

（3）设计加权融合与高风险优先约束相结合的可解释判定机制，并将结果映射为放行、告警、阻断和外部策略联动，形成判定--处置闭环；

（4）在实验环境中开展总体对比、框架级消融和运行可行性测试，验证总体框架的风险判定能力、关键机制贡献及运行开销。

\section{相关工作}

\subsection{Web应用对象级越权访问}

对象级越权通常与IDOR（Insecure Direct Object Reference）和BOLA（Broken Object Level Authorization）相关。OWASP Web Security Testing Guide将IDOR定义为应用直接使用用户可控输入指向数据库记录、文件或其他资源，导致攻击者可以通过修改参数绕过授权检查\cite{owasp-idor}。OWASP Top 10也长期将访问控制失效列为Web应用最重要的安全风险之一\cite{owasp-top10-2025}。

对象级越权并非简单的认证失败。用户可能已经通过登录认证，令牌也处于有效期内，但后端未校验该用户与目标对象之间的授权关系。该类风险既可能发生在读取操作中，也可能发生在修改、删除、审批、导出等具有更高危害的操作中。与传统漏洞相比，对象级越权更依赖业务数据结构和访问上下文，自动化检测难度更高。

\subsection{规则防护与自动化扫描方法}

基于规则的Web防护方法通常通过URL黑白名单、参数约束、敏感路径匹配和签名检测完成过滤。WAF和API网关可快速拦截已知攻击模式，但其策略维护依赖人工配置，难以自动理解“用户A是否有权访问对象B”。当业务接口频繁变化或对象字段命名不统一时，规则方法容易出现漏配和误报。

自动化黑盒扫描工具能够发现部分输入校验、注入和配置类漏洞。Bau等\cite{bau2010}和Doupé等\cite{doupe2010}对黑盒Web漏洞扫描工具进行了系统评估，指出该类工具在状态保持、认证后路径覆盖和业务逻辑漏洞识别方面仍存在局限。RESTler通过API规范推断请求间的生产者--消费者依赖并生成有状态请求序列\cite{restler2019}，增强了复杂REST API的状态覆盖，但其主要目标是测试阶段的缺陷发现，而非生产流量中的在线授权判定。对象级越权通常需要比较不同用户、不同对象和不同角色下的响应与状态，因此仅依靠通用扫描器或接口序列生成仍难以恢复完整授权语义。

\subsection{访问控制模型与运行时监测}

RBAC将权限与角色关联，能够降低权限管理复杂度\cite{sandhu1996}；ABAC进一步将主体、对象、操作和环境等属性纳入授权决策\cite{hu2014}。这些访问控制模型为系统内部权限设计提供了理论基础，但在运行时外部流量分析场景中，检测系统通常无法直接读取业务系统的完整权限配置、数据库关系和代码逻辑。Sun等\cite{sun2011}利用静态分析推断不同角色的访问控制假设，AuthScope则通过差分流量分析与字段替换发现在线服务中的授权缺陷\cite{authscope2017}。前者依赖源代码，后者面向主动测试，两者均说明跨角色或跨主体对照是识别授权缺陷的重要依据。

运行时监测方法尝试通过应用状态、请求序列和历史行为识别异常。Cova等\cite{cova2007}提出基于应用内部状态的异常检测方法，Felmetsger等\cite{felmetsger2010}从逻辑漏洞角度探索Web应用运行时检测。上述研究说明，业务逻辑类安全问题需要结合状态和上下文进行判断。本文与现有静态分析、主动扫描和应用内插桩方法的区别在于：本文以被动或代理侧请求--响应事务为输入，研究如何在流量侧组织多类运行时授权风险证据，并将风险判断直接映射为分级处置。本文关注总体框架与协同机制，各类证据分量仅作为统一决策链中的组成部分进行实现与验证。

\section{方法总体框架}

\subsection{总体架构}

本文框架由流量接入、事务关联、上下文提取、运行时证据组织、风险融合、联动处置和状态审计等部分构成，如图\ref{zh:fig:architecture}所示。系统可部署于反向代理、API网关或旁路观测节点。在线模式在请求转发前基于当前请求上下文和既有运行时证据完成风险判定，并在响应返回后更新事务状态；旁路模式则主要输出审计与告警结果。本文所称“非侵入式”是指无需修改被保护应用的业务代码和内部授权逻辑；反向代理和API网关模式仍可能位于业务流量的在线转发链路中。

\begin{figure}[htbp]
  \centering
  \begin{tikzpicture}[
    x=1cm,y=1cm,
    flowbox/.style={draw=black!70,fill=black!3,rounded corners=2pt,align=center,
      minimum height=0.9cm,inner sep=5pt,font=\small,line width=0.65pt},
    keybox/.style={flowbox,fill=black!9},
    external/.style={flowbox,fill=white,dashed},
    storage/.style={flowbox,double,double distance=0.8pt,fill=black!2},
    flowline/.style={draw=black!65,line width=0.7pt},
    flowarrow/.style={flowline,-{Latex[length=2.3mm,width=1.6mm]}},
    support/.style={draw=black!55,dashed,line width=0.65pt,<->}
  ]
    \node[external,minimum width=2.6cm] (client) at (-4.8,0) {客户端/调用方};
    \node[keybox,minimum width=4.2cm] (ingress) at (0,0)
      {非侵入式流量接入层\\反向代理/API网关/旁路流量};
    \node[external,minimum width=3.2cm] (protected) at (4.8,0)
      {被保护Web应用\\或API系统};

    \node[flowbox,minimum width=3.2cm] (correlation) at (0,-1.55)
      {请求--响应事务关联};

    \node[flowbox,minimum width=3.0cm] (identity) at (-3.6,-3.2)
      {主体与认证上下文};
    \node[flowbox,minimum width=3.0cm] (endpoint) at (0,-3.2)
      {端点与对象上下文};
    \node[flowbox,minimum width=3.0cm] (behavior) at (3.6,-3.2)
      {行为与网络上下文};

    \node[flowbox,minimum width=3.8cm] (evidence) at (0,-4.9)
      {运行时授权风险证据};
    \node[keybox,minimum width=3.8cm] (fusion) at (0,-6.35)
      {风险融合与优先约束};
    \node[flowbox,minimum width=3.7cm] (response) at (0,-7.8)
      {联动处置与审计输出};

    \node[storage,minimum width=3.6cm] (state) at (4.8,-6.35)
      {共享状态存储\\事务、上下文、证据\\策略与审计日志};

    \draw[flowarrow] (client) -- (ingress);
    \draw[flowarrow] (ingress) -- (protected);
    \draw[flowarrow] (ingress) -- (correlation);

    \draw[flowline] (correlation.south) -- (0,-2.35);
    \draw[flowline] (-3.6,-2.35) -- (3.6,-2.35);
    \draw[flowarrow] (-3.6,-2.35) -- (identity.north);
    \draw[flowarrow] (0,-2.35) -- (endpoint.north);
    \draw[flowarrow] (3.6,-2.35) -- (behavior.north);

    \draw[flowline] (identity.south) -- (-3.6,-4.05);
    \draw[flowline] (endpoint.south) -- (0,-4.05);
    \draw[flowline] (behavior.south) -- (3.6,-4.05);
    \draw[flowline] (-3.6,-4.05) -- (3.6,-4.05);
    \draw[flowarrow] (0,-4.05) -- (evidence.north);

    \draw[flowarrow] (evidence) -- (fusion);
    \draw[flowarrow] (fusion) -- (response);
    \draw[support] (fusion.east) -- (state.west);
  \end{tikzpicture}%
  \caption{越权风险识别与联动处置总体框架}
  \label{zh:fig:architecture}
\end{figure}
\FloatBarrier

\subsection{主要符号与数据对象}

本文以一次HTTP请求作为在线风险判定单元，并在响应返回后形成关联事务$T$，用于运行时证据更新与审计。表\ref{zh:tab:notation}列出了框架使用的主要数据对象。UserKey、EndpointKey和ObjectId分别用于表示访问主体、归一化端点和业务资源实例；$\mathcal{E}_{u,r}(t)$表示在时刻$t$前，主体$u$对资源类型$r$已形成的运行时对象访问证据集合。

\begin{table}[htbp]
  \centering
  \caption{主要符号及含义}
  \label{zh:tab:notation}
  \begin{tabular}{p{2.8cm}p{9.9cm}}
    \toprule
    符号 & 含义 \\
    \midrule
    $T$ & 一次已关联的HTTP请求--响应事务及其时间、网络和业务上下文，用于证据更新与审计 \\
    UserKey & 访问主体标识，可由账号、会话或低可信临时标识表示 \\
    EndpointKey & HTTP方法与归一化路径组合形成的端点标识 \\
    ObjectId & 指向业务资源实例的候选标识 \\
    $\mathcal{E}_{u,r}(t)$ & 时刻$t$前主体$u$对资源类型$r$已形成的基础运行时对象访问证据集合 \\
    \bottomrule
  \end{tabular}
\end{table}

\subsection{数据处理流程与部署模式}

系统接收客户端发往Web应用的HTTP/HTTPS请求后，首先从请求头、Cookie、JWT、查询参数、表单参数和JSON请求体中提取UserKey、EndpointKey、ObjectId及认证状态等上下文，并结合既有对象证据和近期访问序列形成风险分量。风险融合模块在请求转发前输出综合风险等级，处置模块据此执行放行、告警、阻断或向外部策略执行点发送指令。当前请求的风险判定仅使用请求到达时可获得的上下文以及此前事务已经形成的运行时证据。对于已放行的请求，系统在响应返回后将请求与响应关联为事务$T$，补充响应摘要和业务结果，并更新供后续请求使用的运行时证据与审计状态；当前响应不反向参与同一请求的风险判定。

在反向代理或API网关模式下，系统可在TLS终止后的明文HTTP层获取完整事务，并支持在线阻断；在旁路模式下，系统通过镜像流量、网关日志或审计日志进行观测和告警。当无法获得应用层明文时，只能进行元数据级分析，无法保证对象标识和业务结果提取的完整性。

\subsection{原型实现与研究范围}

本文聚焦总体框架及其协同机制。为保证框架级机制能够运行并具备可解释性，原型系统采用确定性规则实现各风险分量。对象访问证据由有效访问主体、成功业务结果、预配置事务语义和当前风险判定共同确认，并以二值集合进行维护；身份与认证风险依据账号、会话、令牌、角色声明及近期认证上下文之间的一致性计算；行为风险根据时间窗口内的对象切换、访问频率和请求序列特征确定；端点操作风险则结合HTTP方法、路径语义和预设业务影响等级进行配置。

本文实验主要验证上述组件能否在统一数据流和决策链中协同运行，关键证据与融合机制是否影响判定结果，以及原型能否在受控环境中完成在线风险决策。因此，本文结论限定于所述实验条件下的框架级机制验证。

\section{关键机制设计}

\subsection{请求--响应关联与上下文提取}

请求--响应关联用于将协议消息恢复为可解释的业务事务。事务$T$包含请求方法、归一化路径、请求头、参数与请求体摘要、响应状态码、响应体摘要、业务成功状态、客户端地址和时间戳等信息。HTTP方法与消息语义遵循RFC 9110\cite{rfc9110}，JWT声明字段参考RFC 7519\cite{rfc7519}。

在代理和网关模式下，可依据连接上下文、请求序号、Trace ID或网关注入标识完成确定性关联；旁路模式可依据五元组、请求特征和时间窗口进行近似关联。对于长连接、异步接口和批量请求，系统记录关联置信度，避免将不完整事务用于高可信证据更新。

UserKey优先使用账号标识，其次使用会话或令牌标识；当上述信息不可见时，可结合源地址、User-Agent和设备特征形成低可信临时标识。EndpointKey通过HTTP方法与归一化路径组合生成，路径中的数字串、UUID和随机标识被替换为占位符。ObjectId则从路径、查询参数、表单或JSON字段中提取，并通过字段语义、取值形态和响应结构过滤明显的页码、时间戳和金额字段。

\subsection{运行时授权风险证据}

框架将异构上下文映射为六类基础风险分量：

（1）对象访问证据风险$x_{\mathrm{obj}}$：反映当前ObjectId是否具备已确认的运行时访问证据；

（2）身份一致性风险$x_{\mathrm{id}}$：反映账号、会话、令牌和角色声明之间是否存在明显冲突；

（3）行为异常风险$x_{\mathrm{beh}}$：反映短时间对象切换、连续编号访问或异常访问序列；

（4）认证上下文风险$x_{\mathrm{auth}}$：反映敏感操作是否存在近期有效认证或权限变更事件作为支撑；

（5）网络环境风险$x_{\mathrm{net}}$：反映地址漂移、异常代理或共享出口等辅助信息；

（6）端点操作风险$s_{\mathrm{ep}}$：反映当前HTTP方法和端点操作的潜在业务影响。

对象访问证据的基础表示如式（\ref{zh:eq:evidence-set}）所示。在当前原型中，只有同时满足主体有效、响应确认业务成功、事务符合预配置确认语义，且请求未被判定为中风险或高风险时，该事务才被视为可信事务。可信事务仅用于更新后续请求所使用的证据，不能改变其自身已经形成的判定结果。该定义为框架级验证提供可执行的证据接口。

\begin{equation}
  \mathcal{E}_{u,r}(t)=
  \left\{o\mid o\text{在时刻}t\text{前由主体}u\text{的可信事务确认}\right\}
  \label{zh:eq:evidence-set}
\end{equation}

若当前对象已存在于$\mathcal{E}_{u,r}(t)$中，则对象证据风险较低；若当前对象尚未出现在该集合中，且同时伴随异常对象切换、敏感操作或身份上下文冲突，则风险提高。为避免共享网络和移动网络用户受到不合理影响，网络环境风险不单独触发阻断。

\subsection{风险融合与高风险优先约束}

各风险分量归一化至$[0,1]$。基础风险采用固定权重求和，权重满足$w_i\geq0$且$\sum_{i=1}^{6}w_i=1$。为避免高危证据在多个低风险分量的加权结果中被稀释，框架设置可解释的高风险优先约束。总风险按式（\ref{zh:eq:risk-total}）计算：

\begin{equation}
\begin{aligned}
  R_{\mathrm{base}}&=
  \min\!\left\{1,\max\!\left[0,\sum_{i=1}^{6}w_i x_i\right]\right\},\\
  R_{\mathrm{total}}&=
  \max\!\left(R_{\mathrm{base}},
  \max\!\left(\{h_j\mid j\in\mathcal{H}\}\cup\{0\}\right)\right),
\end{aligned}
\label{zh:eq:risk-total}
\end{equation}

其中，$\boldsymbol{x}=(x_{\mathrm{obj}},x_{\mathrm{id}},x_{\mathrm{beh}},x_{\mathrm{auth}},x_{\mathrm{net}},s_{\mathrm{ep}})$；$\mathcal{H}$表示被触发的高风险条件集合，$h_j\in[0,1]$表示对应风险下限。内层最大值中加入$0$，用于定义未触发任何高风险条件时的计算结果。高风险条件仅由可解释的组合证据触发，例如“对象证据缺失且操作敏感”或“身份声明冲突且缺少有效认证事件”，而不是由单一网络特征触发。

风险判定流程如图\ref{zh:fig:risk-fusion}所示。

\begin{figure}[htbp]
  \centering
  \begin{tikzpicture}[
    x=1cm,y=1cm,
    flowbox/.style={draw=black!75,fill=black!3,rounded corners=1pt,align=center,
      minimum height=1.0cm,inner sep=5pt,font=\small,line width=0.6pt},
    keybox/.style={flowbox,fill=black!8},
    flowline/.style={draw=black!70,line width=0.65pt},
    flowarrow/.style={flowline,-{Latex[length=2.2mm,width=1.5mm]}}
  ]
    \node[keybox,minimum width=3.0cm] (extract) at (0,0) {提取运行时风险证据};

    \node[flowbox,minimum width=3.0cm] (object) at (-4.8,-2.0)
      {对象访问证据};
    \node[flowbox,minimum width=3.0cm] (identity) at (-1.6,-2.0)
      {身份/认证上下文};
    \node[flowbox,minimum width=3.0cm] (behavior) at (1.6,-2.0)
      {行为/网络上下文};
    \node[flowbox,minimum width=3.0cm] (endpoint) at (4.8,-2.0)
      {端点操作风险};

    \node[keybox,minimum width=4.4cm] (fusion) at (0,-4.1)
      {加权融合与高风险优先约束};
    \node[flowbox,minimum width=2.8cm] (threshold) at (0,-5.7) {风险等级映射};

    \node[flowbox,minimum width=3.3cm] (low) at (-4.2,-7.4) {低风险：放行};
    \node[flowbox,minimum width=3.6cm] (medium) at (0,-7.4) {中风险：告警/观察};
    \node[flowbox,minimum width=3.6cm] (high) at (4.2,-7.4) {高风险：阻断/联动};

    \draw[flowline] (extract.south) -- (0,-1.05);
    \draw[flowline] (-4.8,-1.05) -- (4.8,-1.05);
    \draw[flowarrow] (-4.8,-1.05) -- (object.north);
    \draw[flowarrow] (-1.6,-1.05) -- (identity.north);
    \draw[flowarrow] (1.6,-1.05) -- (behavior.north);
    \draw[flowarrow] (4.8,-1.05) -- (endpoint.north);
    \draw[flowline] (object.south) -- (-4.8,-3.15);
    \draw[flowline] (identity.south) -- (-1.6,-3.15);
    \draw[flowline] (behavior.south) -- (1.6,-3.15);
    \draw[flowline] (endpoint.south) -- (4.8,-3.15);
    \draw[flowline] (-4.8,-3.15) -- (4.8,-3.15);
    \draw[flowarrow] (0,-3.15) -- (fusion.north);
    \draw[flowarrow] (fusion) -- (threshold);
    \draw[flowline] (threshold.south) -- (0,-6.55);
    \draw[flowline] (-4.2,-6.55) -- (4.2,-6.55);
    \draw[flowarrow] (-4.2,-6.55) -- (low.north);
    \draw[flowarrow] (0,-6.55) -- (medium.north);
    \draw[flowarrow] (4.2,-6.55) -- (high.north);
  \end{tikzpicture}%
  \caption{运行时风险证据融合与处置流程}
  \label{zh:fig:risk-fusion}
\end{figure}

\subsection{联动处置机制}

处置模块根据$R_{\mathrm{total}}$和策略阈值将请求划分为低、中、高风险。低风险请求放行并记录审计日志；中风险请求触发告警、会话标记或持续观测策略；高风险请求在在线部署模式下予以阻断，并可向WAF、API网关、零信任网关或安全运营平台输出策略指令。

框架保留故障旁路、策略回滚和白名单机制。当检测模块不可用或上下文不完整时，可降级为观察模式，避免安全组件故障直接造成业务不可用。所有证据、判定结果和处置动作均写入审计记录，支持事后解释与策略复核。

\section{实验设计与结果分析}

\subsection{实验环境与数据来源}

为验证总体框架的机制可行性，本文在实验环境中搭建Web应用测试系统，并实现反向代理式检测系统。预设实验脚本生成可重复的请求序列，实验系统同步记录样本标签、框架判定结果和决策耗时。实验不包含真实用户隐私数据，样本类别比例也不代表生产环境中授权风险事件的自然分布。

样本共2000条，其中正常访问1000条、对象访问风险事件600条、权限与上下文风险事件400条，如表\ref{zh:tab:dataset-composition}所示。对象访问风险事件由跨主体对象标识替换和连续对象探测样本组成；权限与上下文风险事件由超出当前角色权限的操作、身份声明冲突和认证上下文异常样本组成。上述聚合分类用于验证不同证据来源在同一框架中的协同判定能力，不对各细分类别作独立性能比较。

\begin{table}[htbp]
  \centering
  \caption{实验样本构成}
  \label{zh:tab:dataset-composition}
  \begin{tabular}{lrr}
    \toprule
    样本类别 & 数量/条 & 占比/\% \\
    \midrule
    正常访问 & 1000 & 50.0 \\
    对象访问风险事件 & 600 & 30.0 \\
    权限与上下文风险事件 & 400 & 20.0 \\
    \midrule
    合计 & 2000 & 100.0 \\
    \bottomrule
  \end{tabular}
\end{table}

\subsection{对比方法与评价指标}

实验设置四组方法进行对比：

（1）URL规则方法，仅依据admin、delete、role、export和download等路径或关键词进行判定；

（2）频率检测方法，根据单位时间请求次数或不同ObjectId的切换次数判断异常；

（3）规则+频率方法，对前两类策略的判定结果取并集；

（4）本文框架，综合事务关联、对象访问证据、身份与认证上下文、行为异常、网络环境和端点操作风险进行判定。

评价指标包括准确率（Accuracy）、精确率（Precision）、召回率（Recall）、F1值、误报率（FPR）和漏报率（FNR）。其中TP表示授权风险事件被正确判定为风险，FP表示正常访问被误判为风险，TN表示正常访问被正确判定为非风险，FN表示授权风险事件未被判定为风险。各指标按式（\ref{zh:eq:metrics}）计算。

\begin{equation}
\begin{aligned}
  \mathrm{Accuracy}&=\frac{TP+TN}{TP+FP+TN+FN}, &
  \mathrm{Precision}&=\frac{TP}{TP+FP},\\
  \mathrm{Recall}&=\frac{TP}{TP+FN}, &
  F1&=\frac{2\,\mathrm{Precision}\,\mathrm{Recall}}
  {\mathrm{Precision}+\mathrm{Recall}},\\
  \mathrm{FPR}&=\frac{FP}{FP+TN}, &
  \mathrm{FNR}&=\frac{FN}{FN+TP}.
\end{aligned}
\label{zh:eq:metrics}
\end{equation}

\subsection{框架验证结果}

表\ref{zh:tab:overall-results}给出了四组方法在2000条样本上的框架验证结果。URL规则方法将517条授权风险事件判定为风险，同时将231条正常访问误判为风险，F1值为59.15\%；频率检测方法未产生正常访问误判，但仅将552条授权风险事件判定为风险，召回率为55.20\%；规则+频率方法的召回率提高至83.70\%，但仍误判231条正常访问。本文框架将998条授权风险事件判定为风险，正常访问未出现误判，准确率、召回率和F1值分别为99.90\%、99.80\%和99.90\%。

\begin{table}[htbp]
  \centering
  \caption{总体框架验证结果对比}
  \label{zh:tab:overall-results}
  \resizebox{\textwidth}{!}{%
  \begin{tabular}{lrrrrrrrr}
    \toprule
    方法 & TP & FP & TN & FN & 准确率/\% & 精确率/\% & 召回率/\% & F1/\% \\
    \midrule
    URL规则   & 517 & 231 & 769  & 483 & 64.30 & 69.12  & 51.70 & 59.15 \\
    频率检测  & 552 & 0   & 1000 & 448 & 77.60 & 100.00 & 55.20 & 71.13 \\
    规则+频率 & 837 & 231 & 769  & 163 & 80.30 & 78.37  & 83.70 & 80.95 \\
    本文框架  & 998 & 0   & 1000 & 2   & 99.90 & 100.00 & 99.80 & 99.90 \\
    \bottomrule
  \end{tabular}}
\end{table}

图\ref{zh:fig:overall-metrics}直观比较了四种方法在各项评价指标上的结果。

\begin{figure}[htbp]
  \centering
  \begin{tikzpicture}
    \begin{axis}[
      width=0.96\textwidth,height=7.2cm,
      ybar,bar width=6pt,
      ymin=0,ymax=105,
      ylabel={指标值/\%},
      symbolic x coords={URL规则,频率检测,规则+频率,本文框架},
      xtick=data,
      x tick label style={font=\small,align=center},
      ymajorgrids=true,grid style={black!15},
      legend style={at={(0.5,-0.20)},anchor=north,legend columns=2,font=\footnotesize,
        /tikz/every even column/.append style={column sep=0.8cm}},
      legend image code/.code={\draw[#1] (0cm,-0.10cm) rectangle (0.34cm,0.10cm);},
      enlarge x limits=0.18,
      axis line style={black!70},tick style={black!70}
    ]
      \addplot[draw=black,fill=black!15] coordinates {(URL规则,64.30) (频率检测,77.60) (规则+频率,80.30) (本文框架,99.90)};
      \addplot[draw=black,fill=black!35] coordinates {(URL规则,69.12) (频率检测,100.00) (规则+频率,78.37) (本文框架,100.00)};
      \addplot[draw=black,pattern=north east lines,pattern color=black] coordinates {(URL规则,51.70) (频率检测,55.20) (规则+频率,83.70) (本文框架,99.80)};
      \addplot[draw=black,pattern=dots,pattern color=black] coordinates {(URL规则,59.15) (频率检测,71.13) (规则+频率,80.95) (本文框架,99.90)};
      \legend{准确率,精确率,召回率,F1值}
    \end{axis}
  \end{tikzpicture}
  \caption{不同方法的框架验证指标对比}
  \label{zh:fig:overall-metrics}
\end{figure}

实验样本中，URL关键词难以表征对象归属关系的变化，频率策略则难以覆盖低频授权风险事件。本文框架的验证结果来自多类上下文在同一事务中的协同使用。由于测试数据由预设实验脚本生成，相关指标仅用于说明总体框架在所设场景下的判定可行性，不能外推为生产系统中的普遍性能。

\subsection{框架级消融实验}

为验证总体框架中关键机制的贡献，本文设置三组框架级消融，用于分别考察关键证据类别和优先约束机制对判定结果的影响。结果见表\ref{zh:tab:ablation}。去除运行时对象访问证据后，召回率由99.80\%降至68.50\%，说明对象访问证据是识别授权风险的重要基础；去除行为序列证据后，F1值为99.70\%，表明该证据主要补充连续对象切换和慢速探测；去除高风险优先约束后，召回率降至58.60\%，说明仅采用线性加权容易使高危证据在其他低风险分量的加权结果中被稀释。

\begin{table}[htbp]
  \centering
  \caption{框架级消融实验结果}
  \label{zh:tab:ablation}
  \begin{tabular}{lrrrr}
    \toprule
    消融组 & 准确率/\% & 召回率/\% & F1/\% & 漏报率/\% \\
    \midrule
    完整框架 & 99.90 & 99.80 & 99.90 & 0.20 \\
    去除对象访问证据 & 84.25 & 68.50 & 81.31 & 31.50 \\
    去除行为序列证据 & 99.70 & 99.40 & 99.70 & 0.60 \\
    去除高风险优先约束 & 79.30 & 58.60 & 73.90 & 41.40 \\
    \bottomrule
  \end{tabular}
\end{table}

本文所称“对象访问证据”指式（\ref{zh:eq:evidence-set}）所示的基础二值集合，消融结果用于说明该类证据在当前框架级实现中的作用。

\subsection{运行可行性}

为验证框架在代理链路中的运行可行性，本文在并发数为100、请求总数为1000的条件下开展单机测试，结果见表\ref{zh:tab:performance}。决策模块平均计算时延为0.077\,ms，P99时延为0.137\,ms，端到端吞吐量为91.924\,RPS。平均计算时延仅统计风险证据读取与决策过程，吞吐量则包含代理转发、实验应用处理、客户端调度和排队等待，两者不可直接换算。

\begin{table}[htbp]
  \centering
  \caption{单机运行可行性测试}
  \label{zh:tab:performance}
  \begin{tabular}{rrrrrr}
    \toprule
    并发数 & 请求数 & 吞吐量/RPS & 平均决策时延/ms & P99/ms & 峰值内存/MB \\
    \midrule
    100 & 1000 & 91.924 & 0.077 & 0.137 & 357.426 \\
    \bottomrule
  \end{tabular}
\end{table}

该结果仅表明当前单机实现能够完成在线风险决策，不能据此推断其他运行规模或长周期、复杂部署条件下的系统性能。

\subsection{实验小结}

在所构造的实验场景中，多源运行时证据框架相较所选URL规则和频率策略取得了更高的召回率与F1值；框架级消融说明对象访问证据和高风险优先约束是当前实现中影响召回率的主要机制；单机测试则验证了在线风险决策的运行可行性。上述结果共同表明，总体框架在所设验证环境中具有可行性。

\section{讨论}

\subsection{框架适用性与作用机制}

本文围绕流量侧多源证据如何形成可执行的授权风险决策，统一定义数据对象、证据接口、风险约束和处置流程。框架验证与消融结果表明，对象访问证据、高风险优先约束以及身份、行为和端点上下文的联合使用共同影响风险判定结果，说明总体设计在所设实验环境中能够形成可执行的识别与处置闭环。

该框架适用于需要增强运行时授权风险审计的Web应用和API系统。对于新建系统，可作为RBAC或ABAC机制的补充监测层；对于存量系统，可在不大规模修改业务代码的情况下提供独立的风险审计能力；对于多网关或多微服务场景，可在统一入口处组织跨端点上下文。框架还可与零信任架构中的持续评估、最小权限和策略执行点协同\cite{rose2020}，将风险结果输出至网关或安全运营平台，形成统一的联动处置机制。

在当前原型中，$\mathcal{E}_{u,r}(t)$采用二值对象证据表示，身份与认证上下文通过确定性一致性规则进行评价，端点操作风险则依据请求语义和预设业务影响等级配置。因此，本文结论主要用于说明统一证据组织、风险融合和联动处置机制的可行性。

\subsection{有效性威胁与方法局限}

本文实验存在四类主要有效性威胁。其一，样本由实验环境中的预设脚本生成且正负样本均衡，与生产环境中授权风险事件发生率较低、业务流量呈长尾分布的情况存在差异；其二，对比基线主要由URL规则和频率策略构成，尚未覆盖静态分析、差分测试和学习型检测等其他代表性方法；其三，证据确认规则、风险权重和判定阈值与实验接口语义相关，迁移到其他系统时需要结合业务特征重新标定；其四，性能数据来自单机实现，尚不能反映长周期运行和复杂部署条件下的运行开销。因此，本文结果应理解为实验环境中的框架验证，而不能直接外推为任意Web应用中的通用性能水平。

本文方法存在三类一般性局限。第一，流量侧分析依赖应用层内容和关键业务字段的可见性，当加密内容不可访问或对象标识高度混淆时，证据完整性可能下降。第二，原型中的证据规则、风险权重、判定阈值和端点配置与实验系统语义相关，迁移到其他应用时需要重新配置和标定。第三，受控的类别平衡样本和单机实现尚不能反映生产流量中的类别不均衡、长期状态变化和复杂部署条件。因此，仍需结合更多异构Web应用和真实网关流量，对框架的跨场景适用性与运行稳定性进行检验。

\section{结论}

本文针对Web应用合法会话下的越权风险难以从流量侧识别并实时处置的问题，提出一种基于非侵入式流量分析的总体框架。该框架以HTTP请求为在线判定单元，并利用关联后的请求--响应事务组织和更新对象访问证据、身份与认证上下文、行为序列、网络环境及端点操作风险信息，通过加权融合与高风险优先约束形成分级处置结果。

在实验环境中构造的类别平衡测试样本上，本文框架在所设评价指标上优于URL规则、频率检测及二者组合基线；框架级消融验证了对象访问证据和高风险优先约束的作用；单机测试表明当前实现能够完成在线风险决策。上述结果表明，总体机制在所设验证环境中具有可行性，其更广泛适用性仍需结合异构Web应用和真实网关流量进一步检验。

\section*{作者贡献声明}

林思齐：提出研究命题，设计总体方法框架，完成论文主要起草与最终修订；

罗宇超、邱博睿：参与系统设计与实现、实验场景构造及授权风险样本设计；

王怡然：参与实验数据整理、结果分析和图表绘制；

朱梓涵：参与相关工作梳理、论文格式整理和文献校对。

\clearpage
\renewcommand{\refname}{References / 参考文献}
\bibliographystyle{gbt7714-numerical}
\bibliography{references}

\end{document}